\documentclass[article,aps,pra,reprint,showkeys,showpacs,twocolumn,groupedaddress,10pt]{revtex4-2}
\usepackage{amsmath}
\usepackage{mathrsfs}
\usepackage{amsfonts}
\usepackage{amssymb}
\usepackage{graphicx}
\usepackage[colorlinks,linkcolor=blue,anchorcolor=blue,urlcolor=blue,citecolor=blue]{hyperref}
\usepackage{eufrak}
\usepackage{multirow}
\usepackage{mathrsfs}
\usepackage{textcomp}
\usepackage{appendix}

\begin{document}

\title{Efficiently simulating the work distribution of multiple identical bosons with boson sampling}

\author{Wen-Qiang Liu}
\author{Zhang-qi Yin}\email{zqyin@bit.edu.cn}
\affiliation{Center for Quantum Technology Research and Key Laboratory of Advanced Optoelectronic Quantum Architecture and Measurements (MOE), School of Physics, Beijing Institute of Technology, Beijing 100081, China}

\date{\today }

\begin{abstract}
Boson sampling has been theoretically proposed and experimentally demonstrated to show quantum computational advantages. However, it still lacks the deep understanding of the practical applications of boson sampling. Here we propose that boson sampling can be used to efficiently simulate the work distribution of multiple identical bosons. We link the work distribution to boson sampling and numerically calculate the transition amplitude matrix between the single-boson eigenstates in a one-dimensional quantum piston system, and then map the matrix to a linear optical network of boson sampling. 
The work distribution can be efficiently simulated by the output probabilities of boson sampling using the method of the grouped probability estimation. The scheme requires at most a polynomial number of the samples and the optical elements. Our work opens up a new path towards  the calculation of complex quantum work distribution using only photons and linear optics.
\end{abstract}

\keywords{quantum simulation, quantum work distribution, boson sampling, linear optics}

\pacs{03.67.-a, 03.67.Ac, 03.67.Lx, 42.50.Ex, 05.70.Ce}

\maketitle

\section{Introduction}  \label{sec1}

 Work in nonequilibrium systems is a fundamental research topic and has stimulated many research interests in statistical physics \cite{Nonequilibrium1,Nonequilibrium2,Nonequilibrium3}. Quantum work distribution is a key quantity in the thermodynamic analysis of any quantum system, and determines many important thermodynamic properties, such as the free energy difference \cite{Nonequilibrium1} and nonequilibrium work relation \cite{Nonequilibrium2}. Quantum work distribution in a thermally isolated system can be effectively determined by the beginning-time and end-time energy measurements \cite{two-time}. Previously, there are many explorations on the work distribution in the nonequilibrium quantum system, both theoretically \cite{QHO,QP,work-distribution,work-distribution1,work-distribution4} and experimentally \cite{work-distribution2,work-distribution3,work-distribution5}.  However,  these results mainly focused on the single-particle systems.

In recent years, multi-particle work distribution in nonequilibrium processes has received more and more attentions \cite{many-particle,many-particle0,many-particle1,goold2018role}.
The calculation of work distribution for an identical multi-particle system involves the transition probability between multi-particle eigenstates, which may be formidable difficulty due to the interference influence of these particles \cite{Interference0,Interference1,Interference2}. The transition probability between the eigenstates of multiple identical bosons (fermions) associates with a permanent (determinant) of the corresponding transition amplitude matrix between the single-boson (single-fermion) eigenstate. The determinant can be efficiently calculated on a classical computer, while evaluating the permanent is a so-called `\#P-complete' hard \cite{P-complete,P-complete1}. This implies that the calculation of work distribution with multiple bosons may be classically difficult. 

Quantum boson sampling, a remarkably quantum computational supremacy candidate \cite{quantum-supremacy}, was proposed by Aaronson and Arkhipov in 2011 \cite{Aaronson}. Boson sampling emerges as a powerful paradigm to efficiently solve the output probability distribution of photons in a linear optical network, and provides several practical applications in graph theory \cite{graph1,graph2,bao2023very}, decision and function problems \cite{Decision,Decision1}, quantum chemistry \cite{chemistry1,huh2017vibronic,shen2018quantum,wang2020efficient,chemistry2}, random number generation \cite{shi2022unbiased}, and image encryption \cite{shi2023chaotic}. The early proof-of-principle demonstration of boson sampling has experimentally confirmed that the sampling result is related to the matrix permanent \cite{QBS1,QBS2,QBS3,QBS4,QBS5,QBS6}. Recently, boson samplings with tens \cite{QBS7,supremacy} or even more than a hundred of photons \cite{Interofeter} have been experimentally realized, which fully shows a quantum computational advantage over the classical computer. Besides, the scalable implementations of boson sampling utilizing photonic modes of trapped ions have also been proposed \cite{Ions1,Ions2}. Note that when the  boson number is  $N \leq 30$ and mode number is $\mathcal{O}(N^2)$,  certainly a classical computer could simulate the output probability of boson sampling. But when $N$ achieves approximately 50 or larger, and mode number is  approximately $\mathcal{O}(N^2)$, it would completely beyond the capability of the classical computers  \cite{Aaronson}. 

Though the output probability of boson sampling is associated with the permanent of a matrix, it is completely different from the problem of predicting or estimating the permanent via boson sampling. In fact, it is infeasible to directly estimate the individual output probability using boson sampling, as the detected probability is exponentially small and one has to collect exponentially many samples to achieve a reasonable accuracy \cite{Aaronson}. This is one of the main obstacles for limiting the practical applications of boson sampling. Fortunately, it is found that if these output probabilities are grouped and their sums are estimated, the polynomial samples rather than exponential ones are  required for solving the certain problems \cite{chemistry1,huh2017vibronic,shen2018quantum,wang2020efficient,oh2022quantum}. In this way, boson sampling could be a potentially effective method to solve some practical problems.

In this paper, we investigate how to use a boson sampling system to simulate the work distribution of multiple non-interaction identical bosons in a one-dimensional quantum piston. We firstly present a general theory for the work distribution of multiple bosons and establish a connection between the work distribution and boson sampling. Specifically, we numerically calculate the transition amplitude matrix of the  single-boson  eigenstates and program it into an optical network of boson sampling. The work distribution can be efficiently simulated by the output probabilities of boson sampling using the method of the grouped probability estimation (GPE) instead of the individual probability estimation (IPE). We also analyze the effect of system parameters on the work distribution  and finally present a feasibility analysis of the scheme in terms of the resource cost. Our proposed way opens up a new possibility for studying the work distribution problem of quantum thermodynamics via the linear quantum optical network of boson sampling. 

\section{Results} \label{Sec2}


\subsection{Work distribution with multiple  bosons}  \label{Sec2.1}

\begin{figure} 
\begin{center}
\includegraphics[width=8.1 cm,angle=0]{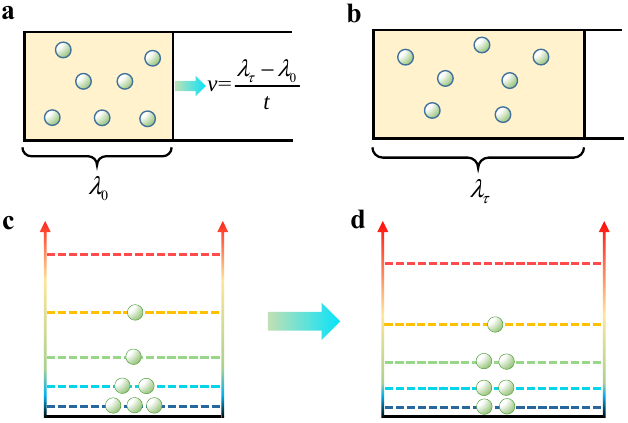}
\caption{Schematic diagram of performing work on the multiple identical bosons  in a one-dimensional quantum piston. \textbf{a} At initial time $t=0$, multiple identical bosons are initially prepared in a thermal state in a keep-temperature quantum piston box with an initial length $\lambda_0$. The piston system then performs work on the bosons by pulling towards the right at a constant speed $v$. \textbf{b} The length of the box becomes $\lambda_\tau$ at final time $t=\tau$. \textbf{c} The initial bosons distribution in the energy eigenstates at $t=0$. \textbf{d} The final bosons distribution in the energy eigenstates at $t=\tau$ after the work being performed.} \label{quantum-piston}
\end{center}
\end{figure}

\begin{figure*}    [htpb]
\begin{center}
\includegraphics[width=17 cm,angle=0]{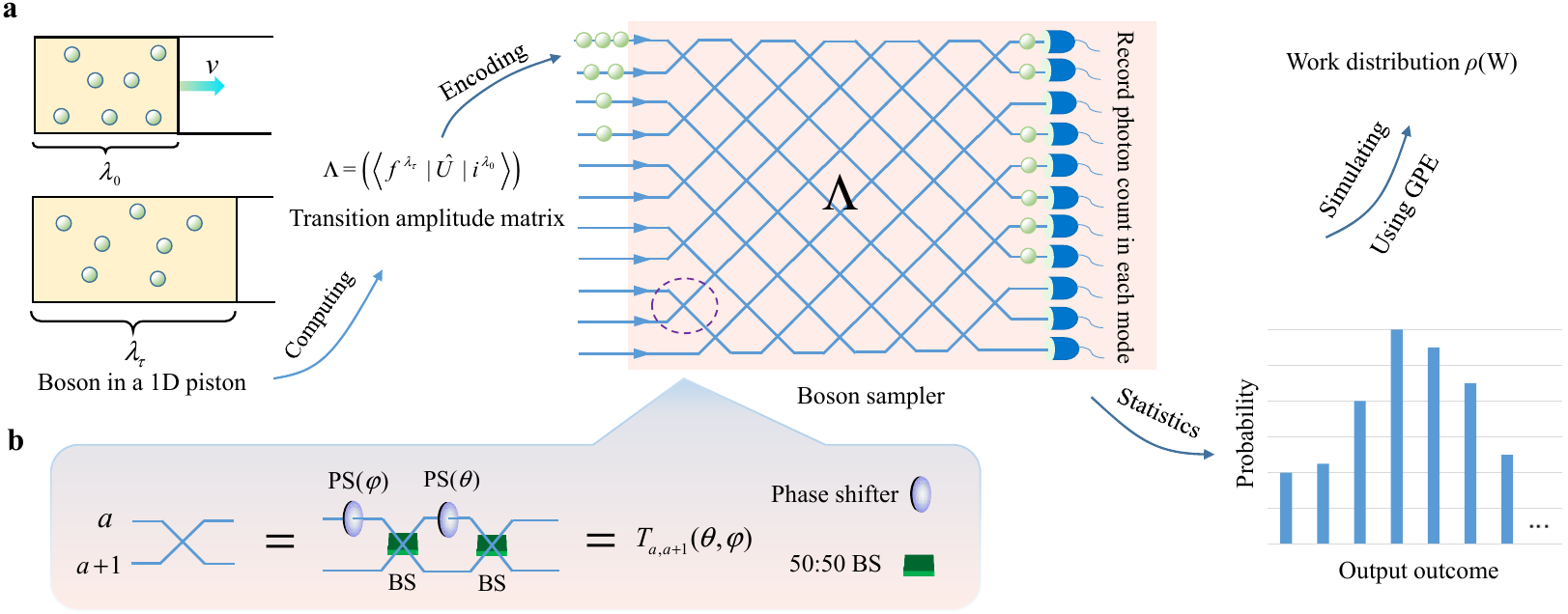}
\caption{Schematic diagram for simulating the work distribution of multiple identical bosons via boson sampling. \textbf{a} The input Fock state injected in the optical modes is used to prepare energy eigenstate of the initial thermal state. The transition amplitude matrix $\Lambda$  between single-boson eigenstates is mapped into an optical network of the boson sampling system. The transition probability between the multi-boson eigenstates is obtained by counting the photon probability distribution and then the work distribution can be simulated by using the method of GPE.  \textbf{b} Variable beam splitter, a key building block for constructing the programmable linear optical network. Each line denotes an optical mode and the crossing line between two modes is a variable beam splitter. The variable beam splitter acting on modes $a$ and $a+1$ is described by a unitary matrix $T_{a,a+1}(\theta,\varphi)$, which can be realized by two 50:50 beam splitters and two phase shifters with the rotated angles $\varphi$ and $\theta$.} \label{sampling}
\end{center}
\end{figure*}

We consider that $N$ identical bosons in a quantum system are driven by a varied external work parameter $\lambda$ from initial time $t=0$ to final time $t=\tau$. The work parameter $\lambda$ could be the position of the quantum piston or the spring coefficient of the harmonic oscillator.  Suppose that at initial time $t=0$, the parameter is $\lambda(0)=\lambda_0$ and the system is prepared in a thermal state with a heat bath at an inverse temperature $\beta=1/(k_BT)$.  Here $k_B$ is the Boltzmann constant and $T$ is the temperature of the system. After preparing a thermal state, we perform a projective measurement over the energy eigenstates of the system. The probability of bosons initially occupying in the energy eigenstate $|i^{\lambda_0}: n_{i}^{\lambda_0}\rangle$ is expressed by 
\begin{align}         \label{eq2}
P(|i^{\lambda_0}: n_{i}^{\lambda_0}\rangle)=\frac{1}{Z^{\lambda_0}}\text{exp}\big({-\beta n_{i}^{\lambda_0} E_{i}^{\lambda_0}}\big). 
\end{align}
Here $Z^{\lambda_0}$ is a partition function, which is given by $Z^{\lambda_0}=\sum_{i}\text{exp}\big({-\beta n_{i}^{\lambda_0} E_{i}^{\lambda_0}}\big)$. $E_{i}^{\lambda_0}$  is the $i$th eigenenergy at initial time $t=0$, and its corresponding eigenstate is $|i^{\lambda_0}\rangle$. $n_{i}^{\lambda_0}$ represents the initial occupation number of bosons in the $i$th eigenstate $|i^{\lambda_0}\rangle$. 

Then the system is detached from the heat bath, and the work parameter of the system is changed from $\lambda(0)=\lambda_0$ at initial time $t=0$ to $\lambda(\tau)=\lambda_\tau$  at final time $t=\tau$ to perform the work on the system. Finally, we apply the second measurement to project the system into its energy eigenstates again. During the process, the system is evolved under the unitary dynamics and the work distribution of the system can be written as \cite{Tasaki,Talkner} 
%
%
\begin{align}
\begin{split}         \label{eq1}
\rho(W)=&\sum_{i}\sum_{f}P\big(|i^{\lambda_0}: n_{i}^{\lambda_0}\rangle\big)  \\&
\times P\big(|i^{\lambda_0}: n_{i}^{\lambda_0}\rangle\rightarrow|f^{\lambda_\tau}: n_{f}^{\lambda_\tau}\rangle\big)  \\&
\times \delta\big(W-n_{f}^{\lambda_\tau}E_{f}^{\lambda_\tau}+n_{i}^{\lambda_0}E_{i}^{\lambda_0}\big). 
\end{split}
\end{align}
Here $E_{f}^{\lambda_\tau}$ is the $f$th eigenenergy at final time $t=\tau$, and its corresponding eigenstate is  $|f^{\lambda_\tau}\rangle$. $n_{f}^{\lambda_\tau}$ represents the final occupation number of bosons in the $f$th eigenstate $|f^{\lambda_\tau}\rangle$. $P\big(|i^{\lambda_0}: n_{i}^{\lambda_0}\rangle\rightarrow|f^{\lambda_\tau}: n_{f}^{\lambda_\tau}\rangle\big)$ is a transition probability from the initial multi-boson eigenstate $|i^{\lambda_0}: n_{i}^{\lambda_0}\rangle$ to the final multi-boson eigenstate $|f^{\lambda_\tau}: n_{f}^{\lambda_\tau}\rangle$.  The work distribution of multiple particles can be further rewritten as 
\begin{align}
\begin{split}         \label{eq2.5}
\rho(W)=\sum_\mathbf{F}P\big(\mathbf{I}) 
\cdot P\big(\mathbf{F}|\mathbf{I}\big)
\cdot  \delta\big(W-n_{f}^{\lambda_\tau}E_{f}^{\lambda_\tau}+n_{i}^{\lambda_0}E_{i}^{\lambda_0}\big). 
\end{split}
\end{align}
Here the probability of initial thermal states is described by a vector $P(\mathbf{I})$ and $P(\mathbf{I})=(P(|1^{\lambda_0}:n_1^{\lambda_0}\rangle), P(|2^{\lambda_0}:n_2^{\lambda_0}\rangle), \cdots)$. $\mathbf{I}=(|1^{\lambda_0}:n_1^{\lambda_0}\rangle, |2^{\lambda_0}:n_2^{\lambda_0}\rangle, \cdots)$ is the  initial eigenstate vector and its corresponding initial boson number distribution is given by $\mathbf{n_{I}^{\lambda_0}}=(n_1^{\lambda_0}, n_2^{\lambda_0}, \cdots)$. $\mathbf{F}=(|1^{\lambda_\tau}:n_1^{\lambda_\tau}\rangle, |2^{\lambda_\tau}:n_2^{\lambda_\tau}\rangle, \cdots)$ is the final eigenstate vector and its corresponding final boson number distribution is given by $\mathbf{n_{F}^{\lambda_\tau}}=(n_1^{\lambda_\tau}, n_2^{\lambda_\tau}, \cdots)$.  $P(\mathbf{F}|\mathbf{I})$ denotes the transition probability from the initial multi-boson eigenstate vector $\mathbf{I}$ to the final multi-boson eigenstate vector $\mathbf{F}$. From Eq. (\ref{eq2.5}), one can see clearly that the work distribution is mainly determined by two factors.  One is the initial thermal distribution probability in Eq. (\ref{eq2}), which can be easily calculated on the classical computer. The other is the transition probability $P(\mathbf{F}|\mathbf{I})$ between the multi-boson eigenstates, which is a classically difficult problem to calculate. Due to the interference of multiple identical bosons, the transition probability between multi-boson eigenstates can be constructed from the permanent of the transition amplitude matrix between single-boson eigenstates and it is expressed by \cite{many-particle,Interference0,Interference1} 
%
\begin{align}
\begin{split}             \label{eq3}
P\big(\mathbf{F}\,|\,\mathbf{I}\big)=\prod_{i=1}\frac{1}{n_{i}^{\lambda_0}!}\prod_{f=1}\frac{1}{n_{f}^{\lambda_\tau}!}\bigg|\text{Per}\big(\Lambda^{(\mathbf{n_{F}^{\lambda_\tau}},\mathbf{n_{I}^{\lambda_0}})})\bigg|^2.
\end{split}
\end{align}
Here matrix function Per($\Lambda$) represents the permanent of a matrix  $\Lambda$. $\Lambda$=$\big(\langle f^{\lambda_\tau}|\hat{U}|i^{\lambda_0}\rangle\big)$ is a transition amplitude matrix between single-boson eigenstates. $\hat{H}(t)$ is the Hamiltonian of the system, and  $\hat{U}$ denotes an evolutionary unitary operator of the system satisfying  the time-dependent Schr\"{o}dinger equation $i\hbar\partial_t\hat{U}(t)=\hat{H}(t)\hat{U}(t)$.  $\Lambda^{(\mathbf{n_{F}^{\lambda_\tau}},\mathbf{n_{I}^{\lambda_0}})}$ denotes a sub-matrix of $\Lambda$ by taking  $n_{f}^{\lambda_\tau}$ ($f=1, 2, \cdots$) copies of the $f$th column and $n_{i}^{\lambda_0}$ ($i=1, 2, \cdots$) copies of the $i$th row of $\Lambda$. Since the total number of bosons is conserved, i.e., $\sum_{i=1}n_i^{\lambda_0}=\sum_{f=1}n_{f}^{\lambda_\tau}=N$, $\Lambda^{(\mathbf{n_{F}^{\lambda_\tau}},\mathbf{n_{I}^{\lambda_0}})}$ occupies a dimension of $N \times N$.

As shown is Fig. \ref{quantum-piston}, $N$ identical bosons in a one-dimensional quantum piston is an interesting example to understand the work distribution. At $t=0$, the bosons are prepared into a thermal state in a stretchable box of length $\lambda_0$ (see Fig. \ref{quantum-piston}\textbf{a}), and their population in the eigenstates is schematically shown in Fig. \ref{quantum-piston}\textbf{c}. Then, the box is stretched to the length $\lambda_\tau$ ($\lambda_\tau>\lambda_0$) at a constant speed $v$ (see Fig. \ref{quantum-piston}\textbf{b}) and at this time the population of bosons in the eigenstates is schematically presented in Fig. \ref{quantum-piston}\textbf{d}. In Appendix \ref{AppendixA},  we provide an explicit expression for constructing  the transition amplitude matrix $\Lambda$ between the single-boson eigenstates \cite{QP,Schrodinger}.  Note that calculating the  transition probability between the multi-boson eigenstates in Eq. (\ref{eq3}) is a classically difficult problem, because the computational complexity of the permanent of a general complex matrix is \#P-hard \cite{P-complete,P-complete1}.  To solve the work distribution problem, in the following text, we would use the boson sampling to simulate the work distribution with multiple identical bosons.

\subsection{Mapping work system into boson sampling} \label{Sec2.2}

Boson sampling is considered as there are $N$ indistinguishable bosons are scattered into a linear unitary network with $M$ optical modes. We denote the input photon state in the Fock basis as $|\text{\textbf{T}}\rangle=|t_1, t_2, \cdots, t_M\rangle$. Each $t_i$ denotes  boson occupation-number in the $i$th optical mode and $|\text{\textbf{T}}\rangle$ describes $N=\sum_{i=1}^{M}t_i$ bosons distribution in each mode. These photons are sent through a linear optical network that is characterized by a unitary transformation $A$. According to the linear mapping relation $a_i^\dagger \rightarrow \sum_{j=1}^{M} A_{ij}b_j^\dagger$ between the input mode creation operator $a^\dagger$ and the output mode creation operator $b^\dagger$ of the network, the probability of getting an output state $|\text{\textbf{S}}\rangle=|s_1, s_2, \cdots, s_M\rangle$ in the Fock basis is mathematically described by \cite{QBS1,QBS2,QBS3,QBS4},
\begin{align}
\begin{split}             \label{eq14}
P\big(\textbf{S}\,|\,\textbf{T}\big)&=\big|\langle \text{\textbf{S}}|A| \text{\textbf{T}}\rangle\big|^2\\&
=\prod_{j=1}^M\frac{1}{s_{j}!}\prod_{i=1}^M\frac{1}{t_{i}!}\bigg|\text{Per}\big(A^\text{(\textbf{S,T})})\bigg|^2.
\end{split}
\end{align}

Remarkably, the probability $P\big(\textbf{S}\,|\,\textbf{T}\big)$ for each of input states and output states is proportional to a permanent of sub-matrix of $A$.  
 Combining Eq. (\ref{eq1}), Eq. (\ref{eq3}), and Eq. (\ref{eq14}), we present a corresponding relationship between the work distribution and boson sampling in Tab. \ref{table1}. As shown in Tab. \ref{table1}, one can see that the space of the work distribution with $N$ bosons and $M$-dimensional transition amplitude matrix is isomorphic to the space of boson sampling with $N$ bosons and $M$-dimensional optical network. Therefore, the work distribution can be obtained by sampling from a great quantity of matrix permanents, equivalently to the boson sampling problem.


%
\begin{table}
\centering\caption{The correspondence between the work distribution and boson sampling.}
\begin{tabular}{llllllll}

\hline  \hline
&Work system        &\quad\;\,&     Boson sampling        \\

\hline

&  Initial thermal state          &\qquad            &  Input Fock state                     \\

&  Amplitude matrix $\Lambda$            &\qquad            & Optical network $A$                      \\

&  Dimension of matrix $\Lambda$         &\qquad                      & Size of network $A$  \\


&  The $i$th energy eigenstate                &\qquad                 & The $i$th mode    \\

&  $\text{Per}\big(\Lambda^{(\mathbf{n_{F}^{\lambda_\tau}}, \mathbf{n_{I}^{\lambda_0}})})$  &\qquad  \quad        &Per\big($A^\text{(\textbf{S,T})}$\big)   \\


&  $P\big(\mathbf{F}\,|\,\mathbf{I}\big)$  &\qquad     & $P\big(\textbf{S}\,|\,\textbf{T}\big)$  \\

\hline  \hline
\end{tabular}\label{table1}
\end{table}

\begin{figure}  [tpb]
\begin{center}
\includegraphics[width=8.5 cm,angle=0]{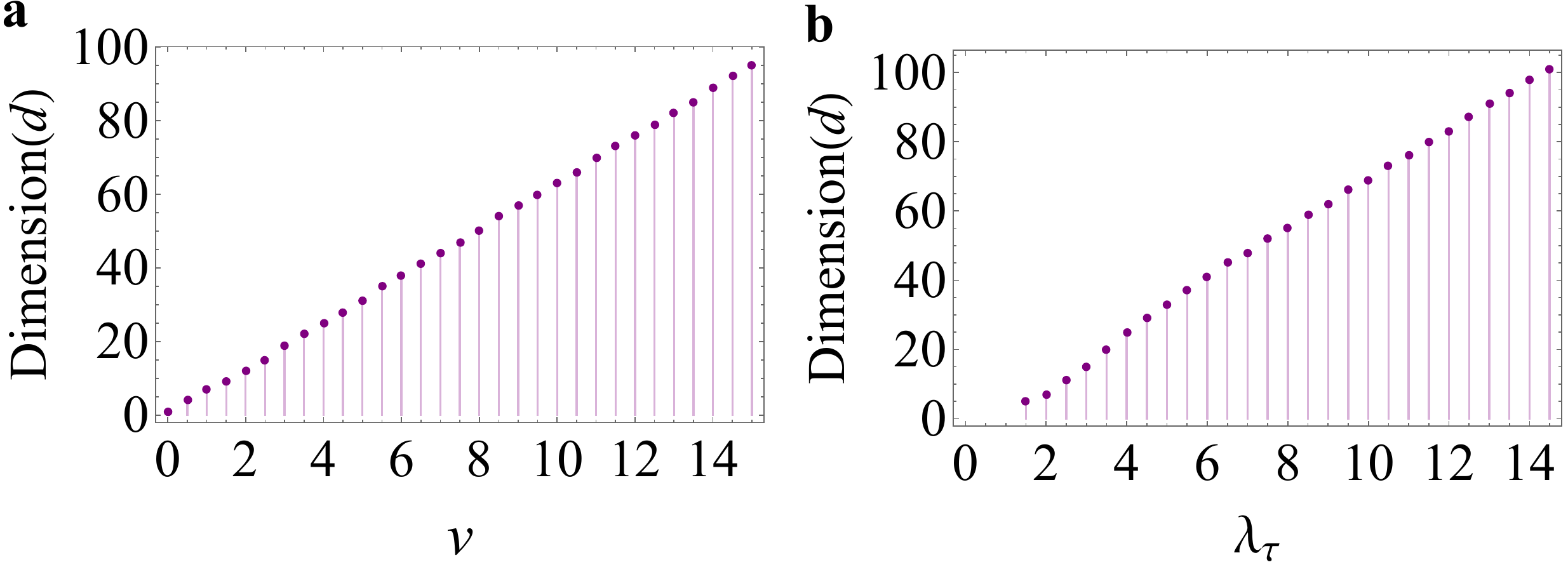}
\caption{\textbf{a} A relationship between the dimension of the matrix $\Lambda$ and the piston expansion speed $v$ when the encoding error is $\mathcal{E}= 0.5\%$. The parameters $\lambda_0=1$ and $\lambda_\tau=2$ are taken. \textbf{b} A relationship between the dimension of the  matrix $\Lambda$  and the final length $\lambda_\tau$ of the box when the encoding error is $\mathcal{E}= 0.5\%$. The parameters $\lambda_0=1$ and $v=1$ are taken.}   \label{dim}
\end{center}
\end{figure}

As shown in Fig. \ref{sampling}, we design a schematic setup for simulating the work distribution of multiple identical bosons via boson sampling.  We first carefully prepare the input Fock state to simulate the initial thermal state of the multiparticle bosonic system.  The input Fock state $|\text{\textbf{T}}\rangle_n$ in the $n$th optical mode represents the $n$th energy eigenstate of the initial thermal state with the probability $P_n$ that can be calculated through the Eq. \eqref{eq2}. The similar method to prepare the thermal state has been experimentally demonstrated \cite{work-distribution5}. We then construct the optical network based on the transition amplitude matrix $\Lambda$ between the single-boson eigenstates. The matrix elements $\langle f^{\lambda_\tau}|\hat{U}|i^{\lambda_0}\rangle$ described in Eq. (\ref{A5}) can be calculated numerically, which depend on the parameters $\lambda_0, \lambda_\tau$, and $v$.  The transition amplitude matrix $\Lambda$ is a unitary matrix in principle because the dimension of the matrix $\Lambda$ could be infinite  and the matrix $\Lambda$ satisfies the normalization, 
\begin{align}
\begin{split}             \label{eq14.5}
\sum_{i}\big|\langle f^{\lambda_\tau}|\hat{U}|i^{\lambda_0}\rangle\big|^2=\sum_{i} \langle f^{\lambda_\tau}|\hat{U}|i^{\lambda_0}\rangle\langle i^{\lambda_0}|\hat{U}^{\dag}|f^{\lambda_\tau}\rangle=1,
\end{split}
\end{align}
\begin{align}
\begin{split}             \label{eq15}
\sum_{f}\big|\langle f^{\lambda_\tau}|\hat{U}|i^{\lambda_0}\rangle\big|^2=\sum_{f} \langle f^{\lambda_\tau}|\hat{U}|i^{\lambda_0}\rangle\langle i^{\lambda_0}|\hat{U}^{\dag}|f^{\lambda_\tau}\rangle=1.
\end{split}
\end{align}
However, in reality we have to restrict the dimension of the matrix  $\Lambda$  to be finite to encode the matrix into a finite dimensional unitary optical network of boson sampling, which causes the matrix  $\Lambda$  to become near-unitary and introduces an encoding error. We can truncate the size of the matrix  $\Lambda$  to make it as unitary as possible to reduce the encoding error. We evaluate the error by calculating the unitary fidelity of the matrix $\Lambda$, which is defined as \cite{Distance-measure} 
\begin{align}
\begin{split}             \label{eq16}
\mathcal{F}=\frac{1}{d}\bigg|tr\sqrt{I_d^{1/2}\sigma I_d^{1/2}}\bigg|.
\end{split}
\end{align}
Here $d$ is the dimension of $\Lambda$ and $I_d$ is a $d$-order identity matrix. $\sigma=\Lambda\Lambda'$ and $\Lambda'$ is the Hermitian conjugate of $\Lambda$.  The encoding error $\mathcal{E}$ is described by $\mathcal{E}=1-\mathcal{F}$. We numerically calculate the matrix elements and truncate the size of matrix $\Lambda$ when the encoding error is 0.5\% with the fixed parameters $\lambda_0=1$, $\lambda_\tau=2$ and a varied speed $v$. A relationship between the matrix dimension and the expansion speed $v$ is plotted in Fig. \ref{dim}a. We also present a relationship between the matrix dimension and the final length of the box in Fig. \ref{dim}b. As shown in Fig. \ref{dim}, one can see that the dimension of the matrix $\Lambda$  increases almost linearly with the acceleration of the piston speed $v$ or the final length $\lambda_\tau$ of the box. Besides, we note that if the truncated $\Lambda$ matrix as a submatrix is embedded into a unitary matrix \cite{bjorklund2018faster} or increasing the dimension of the truncated $\Lambda$ matrix, the encoding error can be further reduced.   In general, the dimension of the optical network will become bigger as the encoding error decreases. 

The next step is to encode the transition amplitude matrix $\Lambda$ into a linear optical network.
There are mainly two configurations of optical network to realize arbitrary unitary matrix, one is the triangle-shaped network \cite{Reck} and the other is the square-shaped network \cite{Clements}. As shown in Fig. \ref{sampling}a, here we use the square-shaped network to realize the transition amplitude matrix, because the symmetry design of the network is more robust against the photon loss and has minimal optical depth and better stability \cite{Clements}.  As shown in Fig. \ref{sampling}b, the crossing between two optical modes $a$ and $a+1$ in the interferometer consists of two 50:50 beam splitters and two phase shifters, which can be expressed mathematically by a matrix $T_{a,a+1}(\theta,\varphi)$ \cite{Clements}. $T_{a,a+1}(\theta,\varphi)$ is called an elimination matrix and it is obtained by replacing the entries of an identity matrix with the same size as $\Lambda$ at the $a$th and $(a+1)$th rows and the $a$th and $(a+1)$th columns with
\begin{align}    \label{eq17}
\left(\begin{array}{cc}
e^{i\varphi}\cos\theta   &  -\sin\theta\\
e^{i\varphi}\sin\theta   &  \cos\theta\\
\end{array}\right),
\end{align}
and the rest of the other entries remain unchanged. Based on Gaussian  elimination method, the matrix $\Lambda$ can be diagonalized into a diagonal matrix $D$ by multiplying a series of $T_{a,a+1}$ and its inverse matrix $T_{a,a+1}^{-1}$. The matrix $\Lambda$ is realized physically in an optical network by choosing suitable values of parameters $\theta$ and $\varphi$ of $T_{a,a+1}$ and the phase values of diagonal matrix $D$ at each output port. The resource overhead for encoding a $d\times d$ unitary matrix into the optical network requires $d(d-1)$ 50:50 beam splitters and $d^2$ phase shifters. In boson sampling system, the phase shifters to realize the diagonal matrix $D$ can be removed as only final photon number is sampled, which will not affect the result but can reduce the resource cost. From Fig. \ref{dim}, one can see that the total resource cost presents at a polynomial hierarchy with the expansion speed $v$ of the piston or the final length $\lambda_\tau$ of the box in terms of the number of required optical elements.

\subsection{The effect of the system parameters on the work distribution}\label{Sec2.3}

\begin{figure}    [tpb]
\begin{center}
\includegraphics[width=8.5 cm,angle=0]{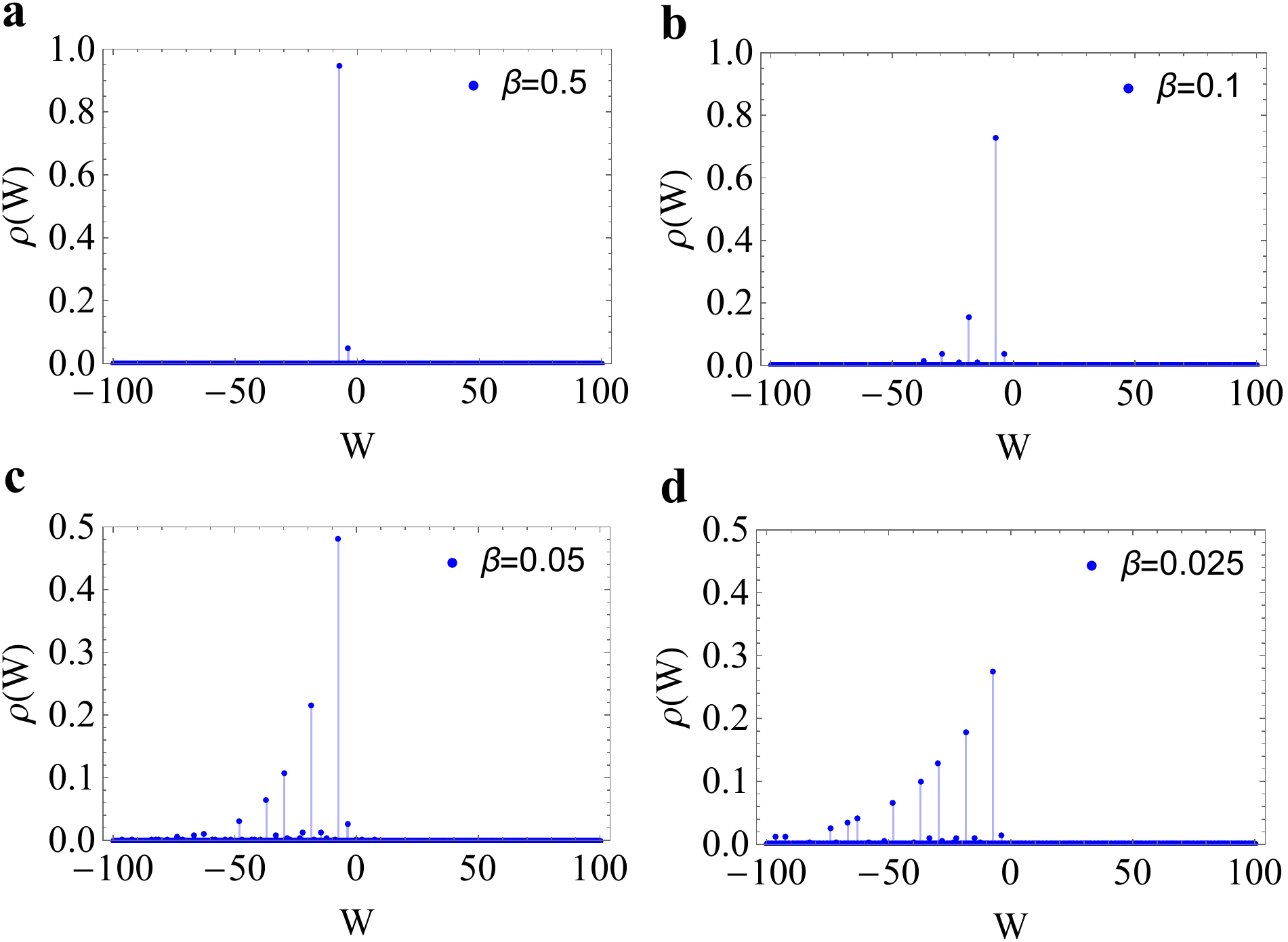}
\caption{The work distribution for two bosons with different temperature in the quantum piston. The temperature rises gradually from $\beta=0.5$ to 0.025, and the system parameters $\lambda_0=1$, $\lambda_\tau=2$, and $v=1$ are taken.} \label{temp}
\end{center}
\end{figure}

The temperature and the speed play an important role in the piston system during the work process.  On one hand, from Eq. (\ref{eq2}), one can see that the temperature of the system affects the initial distribution of the bosons. The work distribution for two bosons with different temperature is presented in Fig. \ref{temp}. As shown in Fig. \ref{temp}, as the temperature increases, the probability that the bosons  populate higher energy levels will increase, which makes the higher energy levels need to be considered. As a result, the dimension of the transition amplitude matrix  will become larger, making the calculation of the work distribution  more complicated.

\begin{figure} [tpb]
\begin{center}
\includegraphics[width=8.5 cm,angle=0]{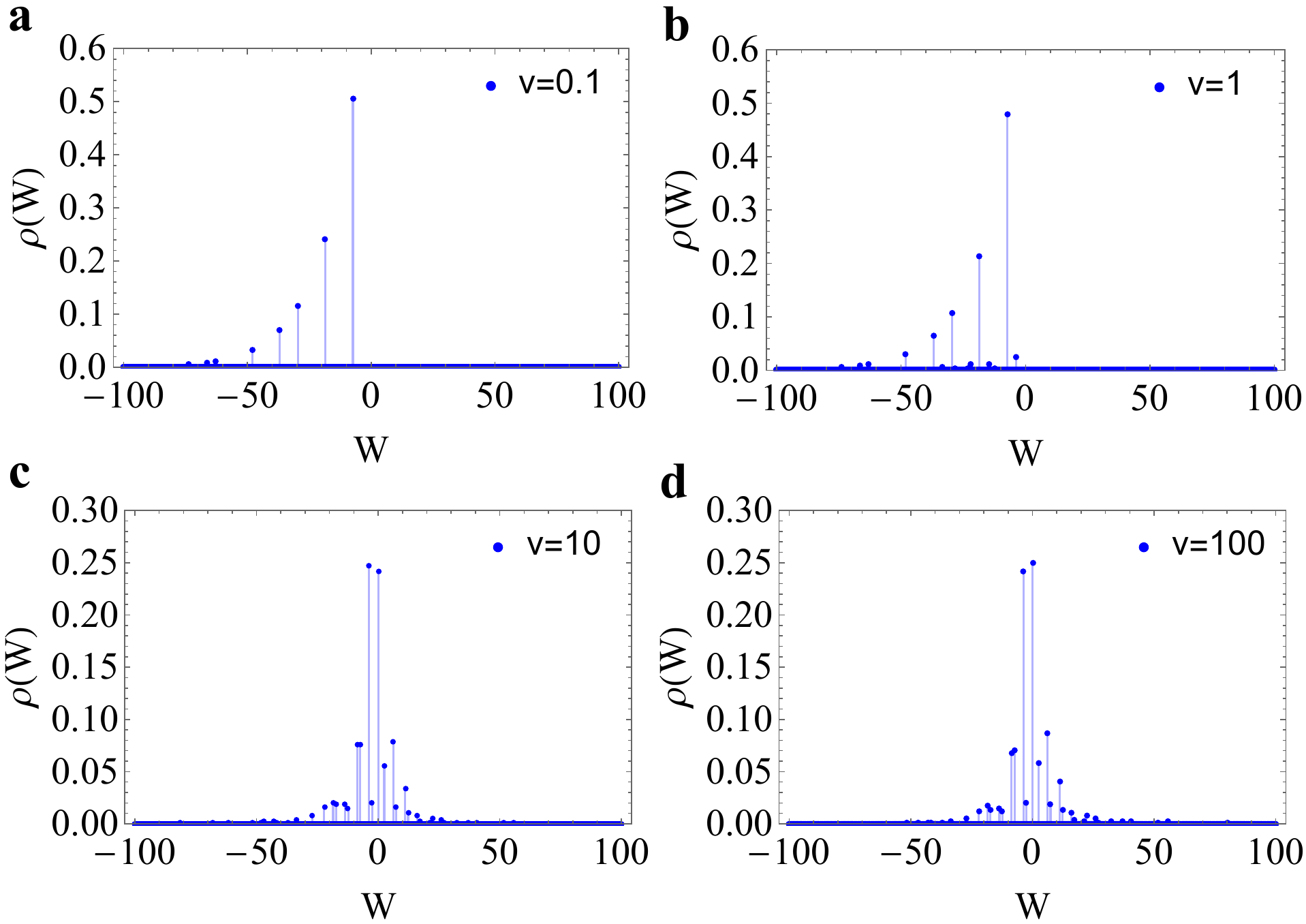}
\caption{The work distribution for two bosons with different moving speed in the quantum piston. The speed rises gradually from $v=0.1$ to 100, and the system parameters $\lambda_0=1$, $\lambda_\tau=2$, and $\beta=0.05$ are taken.} \label{speed}
\end{center}
\end{figure}

On the other hand, the moving speed of the piston is related to the transition probability $P\big(|i^{\lambda_0}: n_{i}^{\lambda_0}\rangle\rightarrow|f^{\lambda_\tau}: n_{f}^{\lambda_\tau}\rangle\big)$. In the low speed limit, one can get the result $P\big(|i^{\lambda_0}: n_{i}^{\lambda_0}\rangle\rightarrow|f^{\lambda_\tau}: n_{f}^{\lambda_\tau}\rangle\big)\rightarrow\delta_{i,f}$  $(v\rightarrow0)$ based on the quantum adiabatic theorem.  Figure \ref{speed}a shows approximately the initial energy distribution: the highest peak represents the energy from the beginning to the end of the ground state, the second highest peak corresponds to the energy of the first excited state, etc. As shown in Fig. \ref{speed}, with the increase of the piston speed, the work distribution for two bosons becomes more complex. This is caused by the higher energy level transition of the bosons when the speed becomes faster.

\subsection{An example}\label{Sec2.4}

\begin{figure}      [tpb]
\begin{center}
\includegraphics[width=6.5 cm,angle=0]{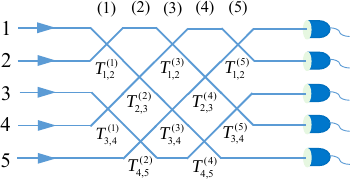}
\caption{Schematic setup for implementing a matrix $\Lambda_{5}$ in an optical network of boson sampling. The numbers (left) are the spatial modes of the  network. The number ($n$) (top) denotes the $n$th time. That is, $T_{a,a+1}^{(n)}$ represents a variable beam splitter acting on spatial modes $a$ and $a+1$ at time $n$. } \label{Lambda_5}
\end{center}
\end{figure}

To understand the work distribution simulated by boson sampling well, as an interesting example, we calculate the work distribution of three bosons in 1D quantum piston in detail.  We consider three bosons are trapped in a box with the initial length $\lambda_0=1$, the stretching speed $v=0.4$, the final length $\lambda_\tau=2$, and the temperature $\beta=0.1$.
We numerically calculate the matrix elements of $\Lambda$ based on Eq. (\ref{A5}) and obtain a $5\times5$ dimensional near-unitary matrix $\Lambda_{5}$ with a   unitary fidelity $\mathcal{F}=0.9992$  (see Appendix \ref{AppendixB}).
We decompose the matrix $\Lambda_{5}$ into the product of a diagonal matrix $D$ and a series of elimination matrices $T_{a,a+1}$ based on Gaussian  elimination method \cite{Clements}. The result of the decomposition is expressed as
\begin{align}
\begin{split}             \label{eq20}
\Lambda_{5}=DT_{3,4}^{(5)}T_{4,5}^{(4)}T_{1,2}^{(5)}T_{2,3}^{(4)}T_{3,4}^{(3)}T_{4,5}^{(2)}T_{1,2}^{(3)}T_{2,3}^{(2)}T_{3,4}^{(1)}T_{1,2}^{(1)}.
\end{split}
\end{align}
As shown in Fig. \ref{Lambda_5}, the matrix $\Lambda_{5}$ is programmed into an optical network of the boson sampler. The values of phase shifter angles $\theta$ and $\varphi$ in the network are calculated and presented in Tab. \ref{table2}. The diagonal matrix $D$ can be ignored without affecting the final output probability of  photons.

\begin{table}[tpb]
\centering \caption{The results of calculating the parameters $\theta$ and $\varphi$ of the elimination matrix $T_{a,a+1}$ for decomposing the matrix $\Lambda_5$. }
\begin{tabular}{cccc}
\hline
\hline
 & Matrix  &\qquad\qquad\qquad\qquad $\theta$   &\qquad\qquad\qquad\qquad $\varphi$ \\
\cline{1-4}
\hline

    & $T_{1,2}^{(1)}$      &\qquad\qquad\qquad \qquad 0.3892    &\qquad\qquad \qquad \qquad 2.9086 \\

    & $T_{3,4}^{(1)}$      &\qquad\qquad \qquad\qquad 0.0852    &\qquad\qquad \qquad\qquad 5.4230   \\

    &  $T_{2,3}^{(2)}$      &\qquad\qquad \qquad\qquad 0.3795    &\qquad \qquad\qquad\qquad 1.9238   \\

    &  $T_{1,2}^{(3)}$      &\qquad \qquad\qquad\qquad 0.0226    &\qquad\qquad\qquad \qquad 3.1496   \\

    &  $T_{4,5}^{(2)}$      &\qquad\qquad \qquad\qquad 0.5999    &\qquad\qquad\qquad \qquad 5.5247  \\

    & $T_{3,4}^{(3)}$      &\qquad\qquad \qquad\qquad 0.0524    &\qquad\qquad \qquad\qquad 3.3019  \\

    &  $T_{2,3}^{(4)}$      &\qquad\qquad \qquad\qquad 0.3493    &\qquad\qquad\qquad \qquad 0.0315  \\

    &  $T_{1,2}^{(5)}$     &\qquad \qquad\qquad\qquad 0.3798    &\qquad\qquad\qquad\qquad 1.8981  \\

    &  $T_{4,5}^{(4)}$      &\qquad\qquad \qquad\qquad 0.6472    &\qquad\qquad \qquad\qquad 3.2375  \\

    & $T_{3,4}^{(5)}$      &\qquad \qquad\qquad\qquad 0.0889    & \qquad\qquad\qquad\qquad 2.5562  \\
\hline\hline
\end{tabular}\label{table2}
\end{table}

\begin{figure} [tpb]
\begin{center}
\includegraphics[width=8.1 cm,angle=0]{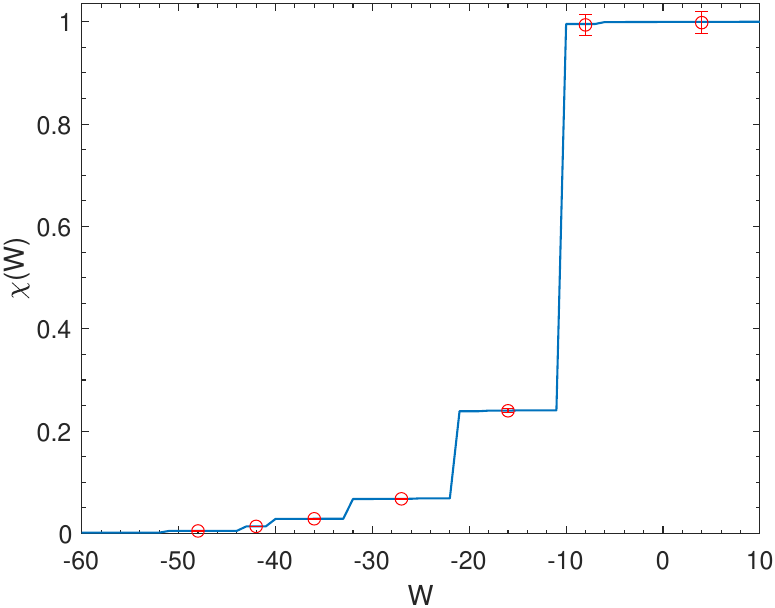}
\caption{The cumulative work distribution for three bosons in an expanding quantum piston system. The parameters $\lambda_0=1$, $\lambda_\tau=2$, $v=0.4$, and $\beta=0.1$ are taken. Error bars represent
the standard deviations in the $100$ numerical simulations with random noise terms $\mathcal{N}$ on the beam splitters. The evaluated work points are (-48, -42, -36, -27, -16, -8, 4) and the corresponding standard deviations are (6.6$\times10^{-5}$, 1.5$\times10^{-4}$, 4.0$\times10^{-4}$, 8.6$\times10^{-4}$, 3.6$\times10^{-3}$, 2.1$\times10^{-2}$, 2.1$\times10^{-2}$). } \label{Work}
\end{center}
\end{figure}

Finally, we calculate the cumulative work distribution by the definition as follows
\begin{align}
\begin{split}             \label{eq21}
\chi(W)=\int^{W}\rho(W')dW'.
\end{split}
\end{align}
The result of cumulative work distribution based on Eq. \eqref{eq21} is plotted in the curve of Fig. \ref{Work}.
In real experiments, the noise and error are inevitable. Therefore, we evaluate the effect of the noise on the cumulative work distribution by adding a random noise $\mathcal{N}\in(-0.01,0.01)$ to the angles of the beam splitters. Explicitly, we randomly choose $100$ groups of noise terms and calculate the cumulative work distribution under the noise effects (see error bars in Fig. \ref{Work}). We find that
the ratio between the error bar and the cumulative work distribution curve is  $1$\% to $2$\%.

\section{Feasibility analysis}\label{Sec3}

Before we discuss the feasibility of the scheme, at first we clarify the relation between boson sampling and the matrix permanent. Boson sampling is a classically difficult problem as the sampled probability involves the permanent of a matrix. However, this does not mean that boson sampling can directly simulate the matrix permanent. Scaling up the system size of boson sampling, the probability of individual output event will become exponentially small, which causes that one has to obtain an exponential number of observations of the event to maintain the accuracy. This method of individual probability estimation is obviously infeasible. Remarkably, if one groups and sums the individual output probabilities so that the sum probability is polynomially small rather than exponentially small, then the number of needed samples would become polynomial size \cite{chemistry1,oh2022quantum}. The GPE can deal with the work distribution problem well.

We next give the explicit method of GPE and evaluate the feasibility of the method by analyzing the required total sample numbers and the  reasonable accuracy.  Based on the delta function in Eq. (\ref{eq2.5}), the output outcomes can be grouped and sum them when $W=m_f E_f^{\lambda_\tau} -n_i E_i^{\lambda_0}$, resulting in,
\begin{align}
\begin{split}             \label{eq24}
G(W)&=\sum_{\textbf{m}}P(\textbf{m}|\textbf{n})\delta\big(W-m_f E_f^{\lambda_\tau}+n_i E_i^{\lambda_0}\big)\\&
=\sum_{\textbf{m}\in\mathcal{G}(W)}P(\textbf{m}|\textbf{n}),
\end{split}
\end{align}
Here $P(\textbf{m}|\textbf{n})$ is the transition probability from the initial photon distribution $\textbf{n}=(n_1, \ldots, n_M)$ to the final distribution $\textbf{m}=(m_1, \ldots, m_M)$. The sets $\mathcal{G}(W)=\{\mathbf{m}\in\mathbb{Z}^M_{\geq0} | m_f E_f^{\lambda_\tau} -n_i E_i^{\lambda_0}=W\}$ for each of grouped outcomes $W\in\{0, \ldots, W_{\text{max}}\}$. To analyse the feasibility of the GPE method, we define two integer energy vectors $\mathbf{E_i}\in \mathbb{Z}^M_{\geq0}$ and $\mathbf{E_f}\in \mathbb{Z}^M_{\geq0}$, in which each of elements $E_{i}$ and $E_{f}$ ($i, f=1, 2, \ldots , M$) are both at most a polynomial large number, i.e., $E_{i}\leq \mathcal{O}\big(\text{poly}(M)\big)$ and $E_{f}\leq \mathcal{O}\big(\text{poly}(M)\big)$. In fact, each initial energy $E^{\lambda_0}_{i}$ and each final energy $E^{\lambda_\tau}_{f}$ in Eq. (\ref{eq2.5}) can be expressed as a floating-point number and they can be transformed as the integer numbers $E_{i}$ and $E_{f}$ by multiplying by a sufficiently large number. Therefore, the number of different groups is $W_{\text{max}}+1\leq \mathcal{O}\big(\text{poly}(M)\big)$, which causes the probability $|G(W)|$ of each group is greater than or equal to $\mathcal{O}\big(1/\text{poly}(M)\big)$ rather than the exponentially small.

Based on the result of grouped probability in Eq. (\ref{eq24}) and the work distribution expression in Eq. (\ref{eq1}), the estimated work distribution can be rewritten as
\begin{align}
\begin{split}         \label{eq25}
\rho_\text{est}(W)=&\sum_{i}P\big(|i^{\lambda_0}: n_{i}^{\lambda_0}\rangle\big)G(W).
\end{split}
\end{align}
This allows us to simulate the work distribution by grouping the output probabilities and collecting at most the polynomial samples of boson sampling.

We next introduce a reasonable accuracy $\epsilon$ to evaluate the performance of the estimated work distribution and the ideal one, i.e.,
\begin{align}
\begin{split}             \label{eq23}
\big|\rho_\text{est}(W)-\rho_\text{ide}(W)\big|\leq\epsilon.
\end{split}
\end{align}
%
Based on the central limit theorem and Chebyshev's inequality, the total sample number $N_\text{tot}$ to achieve a reasonable accuracy scales as Var$(\rho_\text{est}(W))/\epsilon^2$. It is clear that the variance of $\rho_\text{est}(W)$ (denote as Var($\rho_\text{est}(W)$)) is bounded by 1 because the value of the work distribution $\rho_\text{est}(W)$ is between 0 and 1. In other words, $N_\text{tot}=\mathcal{O}(1/\epsilon^2)$ is an upper bound on the total number of samples required to simulate the work distribution, and this upper bound also implies that the reasonable target accuracy becomes  $\mathcal{O}\big(1/\text{poly}(M)\big)$, instead of the exponentially small for individual output probabilities. It suggests that polynomial large samples would be still reasonable.

\section{Conclusion}\label{Sec4}

In conclusion, we have presented a connection between the work distribution and boson sampling. Intuitively, calculating the work distribution may be difficult because the calculation of the transition probability between the multi-boson eigenstates is a classically hard problem.  We found that boson sampling can be used to efficiently simulate the work distribution by sampling the output probability of photons and using the method of the GPE. We analyzed the computational cost with this set-up, and the results showed that at most a polynomial number of observation samples and optical elements are required to achieve a reasonable accuracy.   The connection provides a new possible solution for studying the work distribution that is too difficult to calculate on a classical computer. For other systems, such as multiple bosons in a harmonic oscillator, the calculation of the work distribution also is not an easy problem because calculating the transition probability between the multi-boson eigenstates is necessary \cite{many-particle}. The scheme we developed is  also suitable to  simulate the work distribution of multiple identical bosons in both a contraction piston system and a harmonic oscillator system. 


\section*{ACKNOWLEDGMENTS}

We thank valuable discussions with Zhaohui Wei, Haitao Quan, Xianmin Jin, and Yuanhao Wang.
This work was supported by National Natural Science Foundation of China under Grant No. 61771278 and Beijing Institute of Technology Research Fund Program for Young Scholars.

\appendix

\section*{Appendix}  

\section{Transition amplitudes in a 1D quantum piston} \label{AppendixA}

We show how to obtain analytical solutions to the transition amplitudes $\langle f^{\lambda_\tau}|\hat{U}|i^{\lambda_0}\rangle$ between the
initial energy eigenstates and the final energy eigenstates in a 1D quantum piston  \cite{QP,Schrodinger}. The piston system evolving from time $t=0$ to $t=\tau$ follows the time-dependent Schr\"{o}dinger equation $i\hbar\partial_t\hat{U}(t)=\hat{H}(t)\hat{U}(t)$.  A complete orthogonal solution set of this Schr\"{o}dinger equation can be expressed as \cite{Schrodinger}
\begin{align}
\begin{split}           \label{A1}
\Phi_j(x,t)=\text{exp}\big[\frac{i}{\hbar\lambda(t)}\big(\frac{1}{2}Mvx^{2}-E_j^{\lambda_0}\lambda_0t\big)\big]\phi_j(x,\lambda(t)).
\end{split}
\end{align}
The time-dependent Schr\"{o}dinger equation has the general solution of the following form
\begin{align}
\begin{split}           \label{A2}
\Psi(x,t)=\sum_j^\infty c_j\Phi_j(x,t).
\end{split}
\end{align}
Here, $j=1, 2, \cdots$, and $x$ is the length of the box with the time change, $0\leq x\leq \lambda(t)$. The $j$th eigenenergy $E_{j}^{\lambda_0}$ is given  by $E_{j}^{\lambda_0}=\frac{(j\pi\hbar)^2}{2M{\lambda_0}^2}$, and $M$ is the mass of the boson.
The $j$th eigenstate of a boson in the piston system is
\begin{align}         \label{A3}
\phi_j(x,\lambda)=\sqrt{\frac{2}{\lambda}}\text{sin}\big(\frac{j\pi x}{\lambda}\big).
\end{align}
The coefficients $c_j$ of the solution of Schr\"{o}dinger equation in Eq. (\ref{A2}) can be determined by the initial condition. That is, under the initial condition  $\Psi(x,0)=\phi_{i}(x,\lambda_0)=\langle x|i^{\lambda_0} \rangle$ (taking $M=\hbar=1$), the coefficients become
\begin{align}         \label{A4}
c_j(i)=\frac{2}{\lambda_0}\int_0^{\lambda_0}e^{-i\frac{vx^2}{2\lambda_0}}\text{sin}\big(\frac{j\pi x}{\lambda_0}\big)\text{sin}\big(\frac{i\pi x}{\lambda_0}\big)dx.
\end{align}
The transition amplitudes between the eigenstates of single-particle from time $t=0$ to $t=\tau$ can be expressed as
\begin{align}
\begin{split}             \label{A5}
\langle f^{\lambda_\tau}|\hat{U}|i^{\lambda_0}\rangle=\sum_{j=1}^\infty c_j(i)\int_0^{\lambda_\tau}\Phi_j(x,\tau)\phi_{f}^{*}(x,\lambda_\tau)dx.
\end{split}
\end{align}

\section{Transition amplitude matrix $\Lambda_{5}$} \label{AppendixB}

We numerically calculate the transition amplitudes between the initial energy eigenstates and the final energy eigenstates based on Eq. (\ref{A5}) and obtain a $5\times5$ dimensional near-unitary transition amplitude matrix $\Lambda_{5}$ when the parameters $\lambda_0=1$,  $v=0.4$, and  $\lambda_\tau=2$ are taken. The calculated result of matrix $\Lambda_{5}$ is given by
\begin{widetext}
\begin{eqnarray}    \label{eq19}
\Lambda_{5}=\left(\begin{array}{ccccc}
   0.9843 + 0.1712i   &0.0300 - 0.0273i & -0.0120 - 0.0017i  & 0.0041 + 0.0051i &  0.0003 - 0.0039i\\
  -0.0047 - 0.0401i  & 0.8639 + 0.4990i &  0.0504 - 0.0012i  &-0.0108 - 0.0147i & -0.0017 + 0.0096i\\
   0.0030 + 0.0119i  & 0.0054 - 0.0494i &  0.4535 + 0.8874i  & 0.0394 + 0.0452i  & 0.0070 - 0.0216i\\
  -0.0011 - 0.0069i  &-0.0021 + 0.0186i &  0.0338 - 0.0475i  &-0.3230 + 0.9414i & -0.0232 + 0.0659i\\
  -0.0006 + 0.0044i &  0.0039 - 0.0101i & -0.0166 + 0.0171i  & 0.0662 - 0.0114i & -0.9723 + 0.2054i\\
\end{array}\right).
\end{eqnarray}
\end{widetext}
%

\bibliography{mybibliography}

\end{document}